\documentstyle[psfig,eepic,epic,12pt]{jeet}

\begin{document}
\baselineskip 0.0in
\parindent 0.75in
\parskip 0.0in

\begin{preamble}
 \title{Commensurate and Incommensurate Phases of a Spin-Peierls System } 
 \author{J. Srivatsava} 
 \intphd 
 \submitdate{March 2001}
 \iisclogotrue
 \maketitle
 \acknowledgement

  \hspace{0.75 in}\hspace{2 cm}I wish to express my heartfelt gratitude to my 
guide, Prof. \newline \mbox{Diptiman Sen}, for 
his invaluable guidance and constant support. I am greatly indebted to him for 
his encouragement throughout my course work and project. I also wish to 
thank the faculty 
of CTS and Physics departments, who helped me a great deal in getting the 
requisite 
foundations in physics. 
\vspace{0.3 cm}

\hspace{2 cm}Words wouldn't suffice to express my thanks  to all my batchmates 
and seniors for the help extended by them at various stages of my MS programme. 
I would like to thank my friend, askp, and little sister, 
ammalu, for their 
emotional support.
\vspace{0.3 cm}
  
  \hspace{2 cm}I bow in reverence before God for giving me this great 
opportunity to work in physics. I sincerely wish I can understand many more 
physical phenomena in the years to come.
  
  \vspace{1 cm}
  \hspace{10 cm}
				J. Srivatsava

 \abstract

\hspace{0.75 in} The interest in the study of spin-Peierls phenomenon has
increased with the discovery of copper germanium oxide ($CuGeO_3$) as an example
of spin-Peierls systems. The availability of this inorganic compound as large, 
high quality single crystals, allow experimental measurements which aid in
understanding the spin-Peierls phenomenon. This report details the numerical and
analytical work done on the various magnetic phases that occur in 
a quasi one-dimensional spin-Peierls system (like that of copper germanium
oxide). Taking the system to be finite, the bond displacements, 
single particle energies, energy eigenfunctions, spin densities and  
phonon frequencies are calculated numerically for all possible 
magnetizations, which includes both the  commensurate and 
incommensurate phases, at zero temperature. Solitons of fractional spin 
are found to be present in the system for appropriate magnetizations. 
The uniform to dimerized phase transition occurring in this system is 
also modelled numerically. The single particle and ground-state energies are 
later obtained theoretically for the uniform and dimerized phases. In addition, 
the lattice spin displacements are analytically modelled for a particular
magnetization of the system.

 {
 \baselineskip 1.4pc
 \tableofcontents
 \baselineskip 1.35pc
 \listoffigures
 }
 \baselineskip 1.5pc
\end{preamble}

\parskip      0.35cm
\baselineskip 1.4pc
\chapter{Introduction}

\section{Spin-Peierls systems}

{\Huge A}n interest in the theoretical study of quantum spin systems 
exhibiting 
the spin-Peierls transition has grown ever since the discovery of the 
first 
inorganic spin-Peierls compound, copper germanium oxide ($CuGeO_3$). A 
spin-Peierls transition is a structural transition associated with 
quantum spin 
systems where spin phonon coupling leads to distortions of the lattice. 
This 
magneto-elastic phenomenon gives rise to a variety of phases in the 
system as a 
function of field \textit{H} and temperature \textit{T} \cite{1}. This report is 
aimed at 
analyzing these 
various phases of the spin-Peierls system and the associated physical 
quantities.

In this chapter, the spin-Peierls systems are introduced by first 
discussing the 
structure of the spin-Peierls compound $CuGeO_3$. Its  phase diagram is 
studied 
later and the section is concluded by writing the model Hamiltonian for 
this 
spin-Peierls system.

\section{Structure of $CuGeO_3$}  

\begin{figure}[h]
\centerline{\psfig{figure=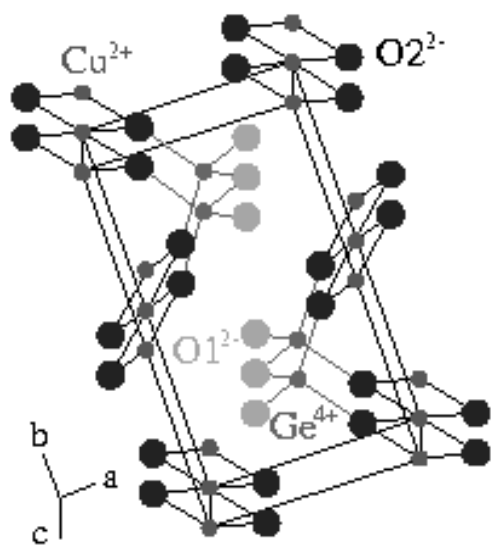,height=8.5 cm}}
\caption{Structure of the spin-Peierls compound $CuGeO_3$}
\label{SF}
\end{figure}

\begin{figure}[!h]
\centerline{\psfig{figure=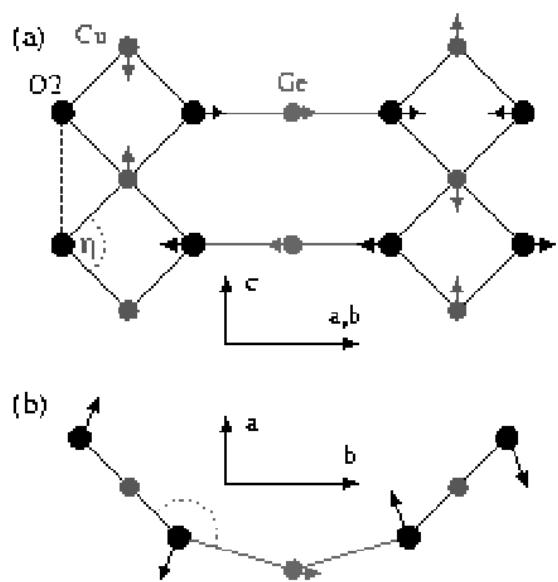,height=8.5 cm}}
\caption{Projection of $CuO_2$ ribbons in (a) three dimensions, and (b) 
a-b plane.}
\label{DF}
\end{figure}

The crystal structure of $CuGeO_3$ is shown in Fig. (\ref{SF}). The 
crystal 
is 
constructed of $CuO_2$ ribbons connected via $GeO_4$ tetrahedra. The 
$Cu^{2+}$ 
ions in $3d^9$ configuration form antiferromagnetic spin 1/2 chain in 
\textit{c} 
direction. There are two different types of $O^{2-}$ shown in the figure 
in
black ($O2$) and grey ($O1$). The $Ge^{4+}$ ions connect adjacent 
$CuO_2$ 
ribbons. The lattice parameters of the orthorhombic unit cell are 
$a=4.81 \AA$,  
$b=8.43 \AA$ and $c=2.95 \AA$ \cite{1}. The Cu-O2-Cu (with a bonding angle $\eta 
\sim 99^o$) 
is responsible for the super exchange intrachain couplings. The lattice 
distortions 
that occur in the dimerized phase obtained by the neutron diffraction 
measurements are shown in Fig. (\ref{DF}). The arrows indicate the 
direction of 
atomic displacements.

\section{(H,T) Phase diagram of $CuGeO_3$} 

\begin{figure}[h]
 \centerline{\psfig{figure=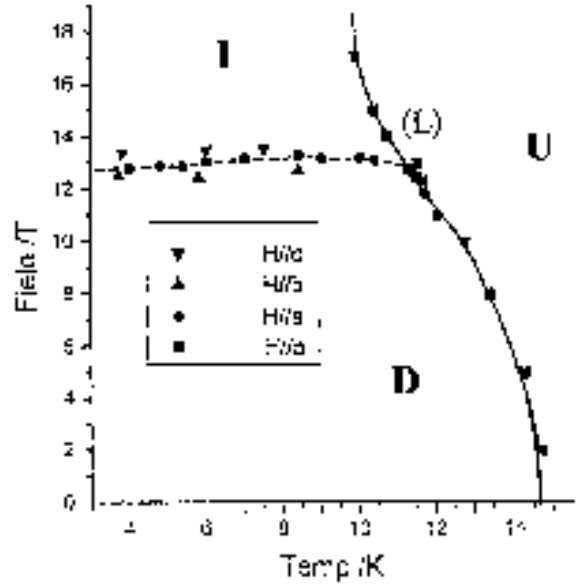,height=8.5 cm}}
\caption{(H,T) phase diagram of the spin-Peierls compound $CuGeO_3$}
\label{PF}
\end{figure}
The (H,T) phase diagram of a spin-Peierls compound $CuGeO_3$ which 
closely 
resembles the theoretical predictions is shown in Fig. (\ref{PF}) \cite{1}. 
The 
diagram 
consists of three phases: the uniform \textit{U} phase, the dimerized D 
phase 
and 
the 
incommensurate \textit{I} phase. At high temperatures, the system is in 
the 
uniform 
phase. In this phase, the magnetic chains have uniform spacing 
\textit{l} and one 
exchange coupling \textit{$J$} between the 
neighbouring spins. As the temperature is decreased, a second order 
structural 
transition is observed at \textit{$T_{sp}$} $\sim$ 14k in the case of 
$CuGeO_3$.  In zero or low fields, this transition occurs and it 
corresponds to a 
dimerization of lattice. In 
that dimerized phase, two lattice parameters \textit{$l_1$} and 
\textit{$l_2$} 
and two 
exchange couplings \textit{$J_{1}$} and \textit{$J_{2}$} are needed to 
define the magnetic 
chains. Increasing the magnetic field from the $D$ phase yields a new 
transition 
occurring at specific value of the field \textit{$H_c$}, $\sim$ 12.5T 
at zero temperature in $CuGeO_3$. This 
field induced transition corresponds to a new deformation of the 
lattice. Above 
\textit{$H_c$}, the lattice becomes incommensurate i.e., the period of 
modulation of spin 
density is generally not a rational multiple of the lattice constant 
\textit{l} as against the 
uniform and dimerized phases. Eventually, as the applied magnetic field 
becomes very large, all the spins get polarized and we get a ferromagnetic 
state.

\section{Modeling the spin-Peierls compound $CuGeO_3$} 

The Hamiltonian $H_{sp}$ used to model the spin-Peierls system is given 
by
\begin{eqnarray}
H_{sp}\hspace{0.2 cm}&=&\hspace{0.2 cm}J\sum_{n=1}^{N} 
\left[1-\lambda\right(u_{n+1} - u_{n} 
\left)\right]\left[S_{n}^xS_{n+1}^x + S_n^yS_{n+1}^y + 
S_n^zS_{n+1}^z\right] \hspace{0.1 cm} \nonumber \\ && \hspace{0.2 cm} 
-h\sum_{n=1}^{N}S_n^z 
\hspace{0.1 cm}  +\hspace{0.1 cm} 
\frac{k}{2}\sum_{n=1}^{N}\left(u_{n+1} - u_{n} \right)^2 \hspace{0.1 
cm} + 
\hspace{0.1 cm}\sum_{n=1}^{N} 
\frac{P_n^2}{2M}
\label{SH}
 \end{eqnarray}

\noindent where \textit{N} is the number of sites, \textit{h} is the 
strength of the magnetic field, ${\vec S}_n$ represents the $\textit{n}^{th}$ 
spin, 
$u_n$ is the displacement of lattice point \textit{n}, $P_n$ represents the 
momentum of lattice point \textit{n}, \textit{M} is the mass of each lattice 
point and $\lambda$ is the positive constant which is 
a measure of the coupling that exists between the lattice points and 
spins. 
In the above equation, periodic boundary conditions are assumed i.e.,  
\begin{math}
\vec{S}_{N+1}\equiv\vec{S_1}.
\end{math}

 The first term in the Hamiltonian gives an antiferromagnetic 
interaction 
between the \textit{nearest 
neighbour} Heisenberg spins along the \textit{c} direction of the 
lattice with 
their 
exchange coupling $J$ modified to take into account the  spin-phonon 
interactions to the first order in bond displacements. Thus  $J$ is 
replaced by $J\left[1-\lambda\right(u_{n+1} - u_{n}  \left)\right]$. 
The 
interactions between the spins in the other two lattice directions and 
between 
the spins other than the nearest neighbour in \textit{c} direction of 
the 
lattice are 
neglected. Considering the distance of separation between the spins, 
this is a 
good approximation and the system can be analyzed as a quasi 
one-dimensional 
(1D) system. The second term is the Zeeman term giving the effect of 
magnetic 
field on the system with net magnetization $\sum_{n=1}^{N}S_n^z$. The 
third term 
is the harmonic energy of the lattice and the final term represents the 
kinetic 
energy of the lattice.
 
 	This Hamiltonian is simplified later to calculate the ground 
state 
\nopagebreak
\enlargethispage*{1000pt}
energies and other physical quantities of interest discussed in the 
next 
chapter. 
 
\chapter{ Numerical Results}
\section{Simplification of the Hamiltonian}
\hspace{0.75in} The full Hamiltonian for the general spin-Peierls 
system is 
shown in the previous chapter, Eq. (\ref{SH}). It is difficult to work 
with 
this Hamiltonian to calculate quantities of interest like single-particle 
energies and wave functions at zero temperature, spin densities  etc. 
Hence the full Hamiltonian is simplified using the
physical arguments discussed below so that the quantities calculated from the 
simplified Hamiltonian reproduce the 
physics of the full Hamiltonian to a very good approximation.

\subsection{Approximations involved and their justification:}
   
\begin{itemize}
	\item{ \textit{Soft phonon limit}: The time scales involved in 
the 
motion 
of lattice are known to be much larger than the time scales 
involved in 
the 
motion of spins arising due to the interactions between the spins and 
phonons. 
Hence the lattice points are initially taken to be at fixed positions 
while 
calculating the unperturbed
single-particle energies at zero temperature and zero field. At the end, the 
lattice 
movements are however taken into account using the perturbation theory 
and the 
perturbed single-particle energies due to motion of lattice points are 
calculated to second order in bond displacements.}
	
	\item{\textit{XY spins}: The generalised Heisenberg spins in the 
model are 
replaced by the XY spins, i.e., the spin-spin interactions in the $\hat{z} $
direction are 
neglected. This is done to simplify the calculations. When the spins are 
transformed into Jordan-Wigner fermions, the interactions in the $\hat{z}$ 
direction involves four spinless fermion operators \cite{4}. It is known from 
numerical 
analysis that reducing the Heisenberg spins 
to XY spins 
does not qualitatively affect the phase diagram of a quasi 1D spin-Peierls 
system.}	
\end{itemize}

The simplified Hamiltonian, $H_s$, thus obtained using the above 
approximations is given by 

\begin{eqnarray}
H_s\hspace{0.2 cm}=\hspace{0.2 cm}-J\sum_{n=1}^{N} 
\left[1+\delta_{n+\frac{1}{2}}\right]\left[S_{n}^xS_{n+1}^x + S_n^yS_{n+1}^y
\right] \hspace{0.1 cm} - \hspace{0.1 cm} h\sum_{n=1}^{N}S_n^z 
\hspace{0.1 cm}+\hspace{0.1 cm} 
\frac{c}{2}\sum_{n=1}^{N}\delta_{n+\frac{1}{2}}^2 
 \label{simpSH}
 \end{eqnarray}
 
 \noindent where $\delta_{n+\frac{1}{2}}=-\lambda\left(u_{n+1} - u_{n} 
\right)$ is the bond displacement and $\lambda$ is the positive constant which 
is a measure of the 
coupling that exists between the lattice points and spins.

It is to be noted that in addition to the mentioned approximations, a unitary 
transformation of the form shown in Eqs. (\ref{T1}) is performed on the 
Hamiltonian given in Eq.(\ref{SH}) to reverse the sign of the first term.

\begin{eqnarray}
S_n^x &\rightarrow &(-1)^nS_n^x \nonumber \\
S_n^y &\rightarrow& (-1)^nS_n^y \nonumber \\
S_n^z &\rightarrow& S_n^z 
\label{T1}
\end{eqnarray}

\subsection{Jordan-Wigner transformation}
Jordan-Wigner transformation transforms a spin Hamiltonian into a spinless 
fermion Hamiltonian \cite{2, 9, 10}. Here, two spinless fermion operators, the 
creation and 
annihilation operators, are defined as shown below

\begin{eqnarray}
C_n &=&\prod_{j=1}^{n-1}(-\sigma_j^z)S_n^- \nonumber \\
C_n^+ &=&\prod_{j=1}^{n-1}S_n^+(-\sigma_j^z)
\label{JW}
 \end{eqnarray}

\noindent where \begin{eqnarray}
S_n^- &=& S_n^x-iS_n^y \nonumber \\
S_n^+ &=& S_n^x+iS_n^y	\nonumber
\end{eqnarray}
\noindent and $\sigma$'s are the Pauli matrices. Using the properties of Pauli 
matrices, it follows from the relations (\ref{JW}) between spin operators and 
fermion operators that the total magnetization,\textit{m}, of the system is

\begin{eqnarray}
m=\sum_{n=1}^{n=N}S_n^z=N_F-\frac{N}{2}
\end{eqnarray}

\noindent where \begin{math}
 N_F=\sum_{n=1}^{N} C_n^+C_n 
 \end{math} is the number of Jordan-Wigner fermions.
 
 Using the Jordan-Wigner transformation on the simplified spin Hamiltonian 
(\ref{simpSH}), a spinless fermion Hamiltonian
is obtained as shown in Eq. (\ref{FH}) 

 \begin{eqnarray}
\tilde{H}\hspace{0.2 cm}&=&\hspace{0.2 
cm}-\frac{J}{2}\left[\sum_{n=1}^{N-1}(1+\delta_{n+\frac{1}{2}})
( C_n^+C_{n+1}+C_{n+1}^+C_n)\hspace{0.1 cm} 
+\hspace{0.1cm}(-1)^{N_F+1}(C_N^+C_{1}+C_{1}^+C_N) \right]
\hspace{0.1 cm} \nonumber \\
&& \hspace{0.2 cm}-  
h\sum_{n=1}^{N}\left[C_n^+C_n-\frac{1}{2}\right]\hspace{0.1 cm}+\hspace{0.1 cm}
\frac{c}{2}\sum_{n=1}^{N}\delta_{n+\frac{1}{2}}^2 
\label{FH}
\end {eqnarray}
 
\section{Single particle energies}

The Hamiltonian used to calculate the single-particle energies at zero 
temperatures for all phases ( for all possible 
magnetizations) is 
given by

 \begin{eqnarray}
\textit{H}\hspace{0.2 cm}&=&\hspace{0.2 
cm}-\frac{J}{2}\sum_{n=1}^{N-1}\left[1+\delta_{n+\frac{1}{2}}\right]\left[C_n^+
C
_{n+
1}+C_{n+1}
^
+C_n\right]\hspace{0.1 cm} 
+\hspace{0.1cm}\left(-1\right)^{N_F}\left[C_N^+C_{1}+C_{1}^+C_N 
\right]
\hspace{0.1 cm}\nonumber \\
& & \hspace{0.2 cm}-  
h\sum_{n=1}^{N}\left[C_n^+C_n-\frac{1}{2}\right]\hspace{0.1 
cm}+\hspace{0.1 
cm} 
\frac{c}{2}\sum_{n=1}^{N}\delta_{n+\frac{1}{2}}^2.
\label{NFH}
\end {eqnarray}

It can be seen that the \textit{boundary term} of this Hamiltonian has 
a 
relative minus sign compared to the Hamiltonian shown in the Eq. 
(\ref{FH}). This appears to break the translation invariance. However, Eq. 
(\ref{NFH})can be made translation invariant by transforming all the fermion 
operators as
  
\begin{eqnarray}
C_n &\rightarrow&  e^{i\pi/N} C_n \nonumber \\
\hspace{0.3cm}C_n^+ &\rightarrow&  e^{-i\pi/N} C_n^+
\end{eqnarray}

\noindent which distributes the minus sign at the boundary over all the sites 
equally.
As $N\rightarrow\infty$, such a transformation produces a vanishingly small 
change in the single-particle energies.

The system size \textit{N} is chosen to be 100 spins. The constant $c$ in the 
final term in the Hamiltonian (\ref{NFH}) showing the harmonic energy of the 
lattice is set equal to unity. The lattice constant 
\textit{l} is also fixed to be unity so that all lengths which are multiples of 
the 
lattice 
constant are now just integers. In order to obtain the 
single-particle energies for any particular 
magnetization or any particular filling of Jordan-Wigner fermions, 
$N_F$, of the 
ground-state, it is necessary to obtain the set of bond 
displacements, $\delta_{n+\frac{1}{2}}$'s which minimize the Hamiltonian given 
in 
Eq. (\ref{NFH}). The procedure followed to achieve this is

\begin{itemize}
\item{Set all $\delta_{n+\frac{1}{2}}$'s to zero initially.}
\item{Increment or decrement each $\delta_{n+\frac{1}{2}}$ separately in steps 
of 
0.0005 and change each of the remaining spins by $0.0005/99$
with the reverse sign, so that for each such change the sum of all 
displacements continues to remain zero as required for a periodic lattice.}
\item{Calculate the new ground-state energy for each such change and iterate in 
a loop running over all the lattice sites, so as to minimize the Hamiltonian.}
 \end{itemize}

\noindent Then the eigenvalues of the Hamiltonian with these bond
displacements, $\delta_{n+\frac{1}{2}}$'s give the single-particle energies for 
the 
particular filling at zero temperature. The eigenfunctions similarly give the 
single-particle wave functions. 
The bond displacements and single-particle energies are obtained for all 
possible magnetizations, few interesting ones of which are shown in the 
following 
sections.

\subsection{For zero magnetization (for $N_F=50$)}

\begin{figure}[h]
 \centerline{\psfig{figure=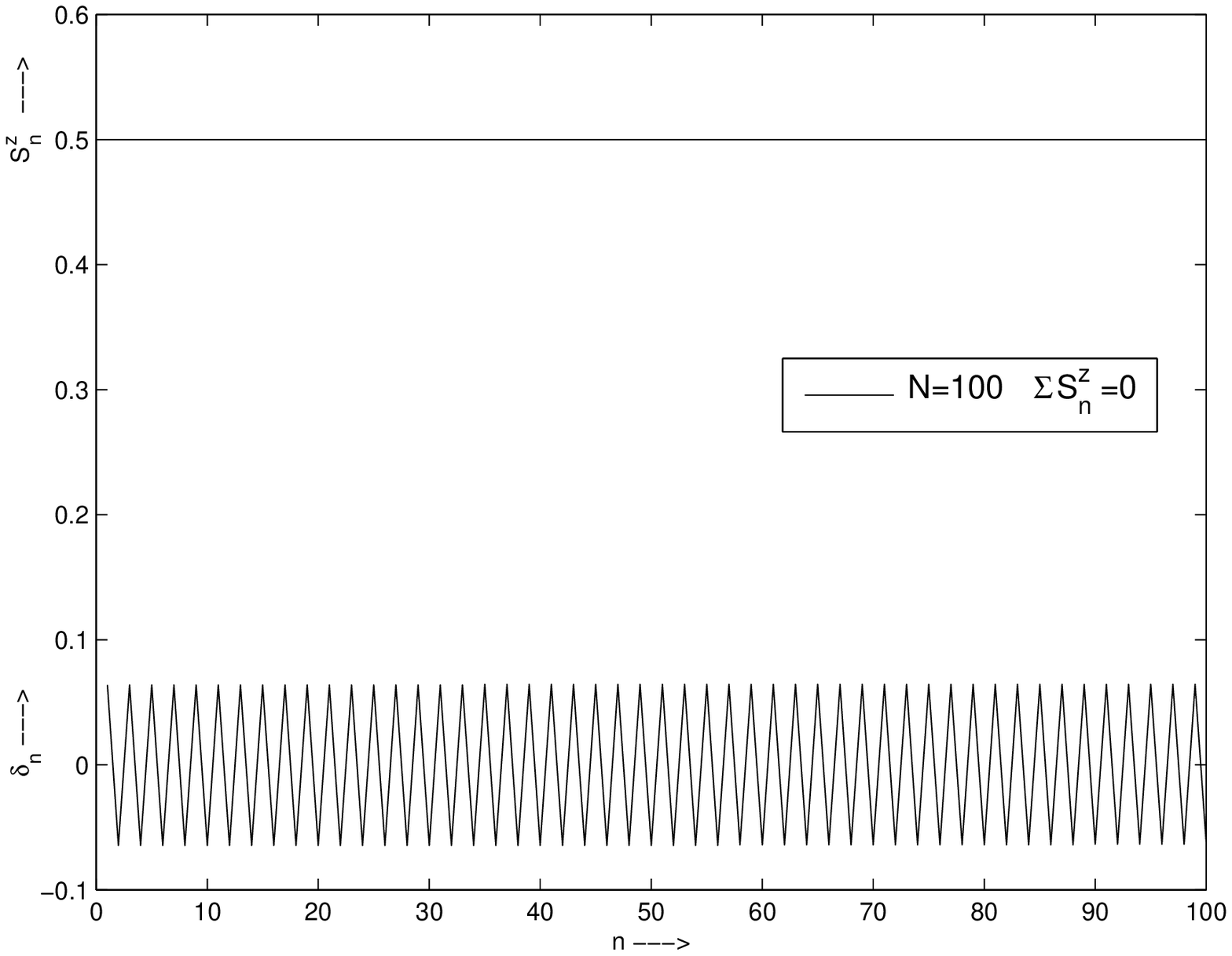,height=10.5 cm}}
 \caption{The bond displacements and expectation values of $S_n^z$ for zero 
magnetization.}
 \label{D50}
 \end{figure}

The bond displacements for the case of zero magnetization, i.e., for 
$N_F$=50, obtained by minimizing the Hamiltonian in Eq. (\ref{NFH}) using the 
above procedure are shown in the lower part of Fig. (\ref{D50}). It can be 
seen that the value of the displacement of the bond alternates in sign 
between any two consecutive lattice spins with its magnitude remaining 
constant. This corresponds to dimerization of the lattice.
The magnitude of each of the bond displacements, $\delta$, is 0.0642.

\begin{figure}[h]
\centerline{\psfig{figure=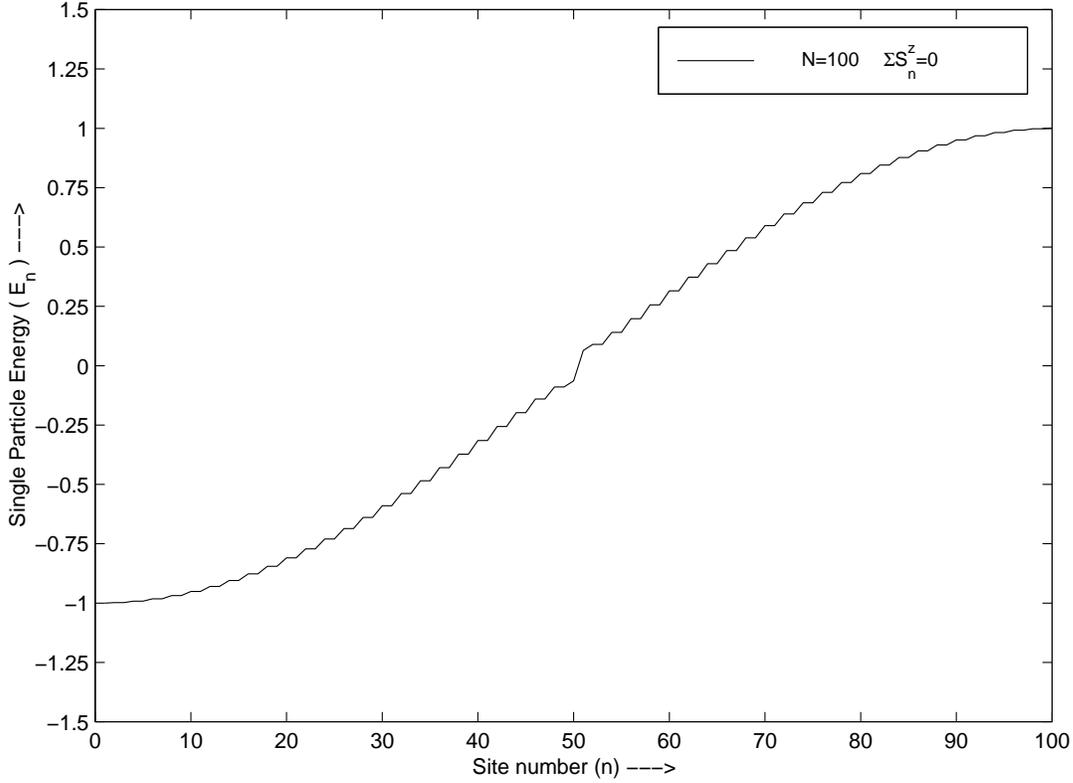,height=10.5 cm}}
\caption{The single-particle energy spectrum of the zero magnetization.}
\label{E50}
\end{figure}

 The single-particle energy spectrum obtained by diagonalizing this Hamiltonian  
with these bond displacements is shown in Fig. (\ref{E50}). It is observed 
that  
 
 \begin{itemize}
 \item{There are an equal number of positive and negative energy levels.}
 \item{For every positive energy level there is a corresponding negative energy 
level which is equal in magnitude but opposite in sign.}
 \item{Each level is doubly degenerate. Exceptions to this fact are the maximum 
level (1), the minimum level (-1), the largest negative level (-0.0642) and the 
smallest positive level (0.0642) which are non-degenerate.}
 \item{There is a \textit{gap} between the largest negative level and 
smallest positive level which is twice of $\delta$. It is to be noted that all 
the quoted energies are in units of $J$.} 
 \end{itemize}

  The ground-state energy of this system is obtained by summing the lowest 
fifty 
levels i.e., all the states lying below the gap (negative energy states). This 
is the state of the system at zero temperature and 
its energy is -31.8634.  
  	
  	The upper half of Fig. (\ref{D50}) gives the variation of spin 
density 
with the site index. It can be seen that this remains constant at half 
throughout the chain as expected. Also the bond displacements, single-
particle 
energies and spin densities remain independent of an applied magnetic field for 
small values of the field because the 
net magnetization in this case is zero.

\subsection{For magnetization=1 (for $N_F$=51)}

 \begin{figure}[!h]
 \centerline{\psfig{figure=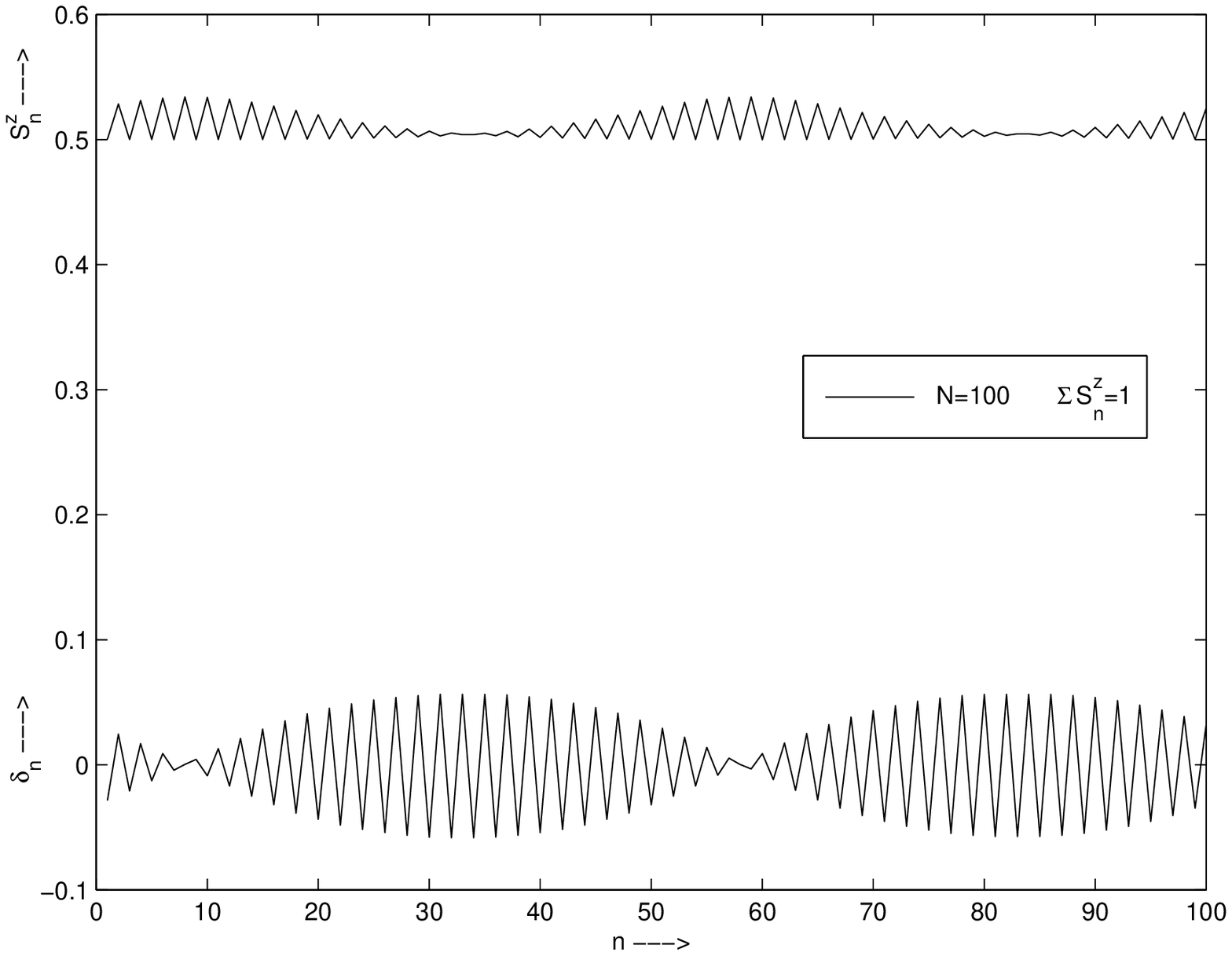,height=10.5 cm}}
 \caption{The bond displacements and expectation values of $S_n^z$ for 
magnetization=1.}
 \label{D51}
\end{figure}
 
 \begin{figure}[!h]
\centerline{\psfig{figure=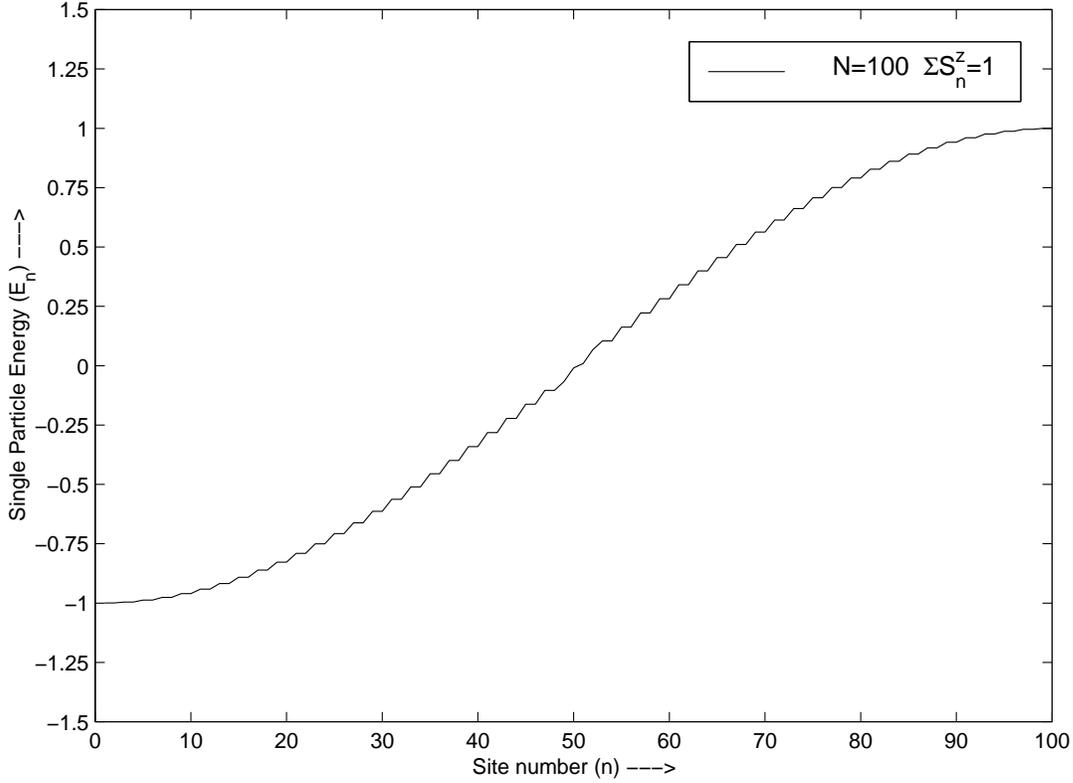,height=10.5 cm}}
 \caption{The single-particle energy spectrum for magnetization=1.}
\label{E51}
\end{figure}

The bond displacements for the case of 51 fermions i.e., one away from 
dimerization, obtained by minimizing the Hamiltonian in Eq. 
(\ref{NFH}) for the case of zero magnetic field using the same procedure, are 
shown in the lower part of Fig. (\ref{D51}). It can be seen that the 
displacements of the bonds oscillate around the value zero and have an 
envelope whose 
periodicity is 50, their maximum magnitude being 0.058. The pattern obtained is 
similar to that of beats in 
sound waves. 

 The single-particle energy spectrum obtained in this case is shown in Fig. 
(\ref{E51}). The features of the spectrum that get modified compared to 
$N_F=50$ 
are 
 
 \begin{itemize}
 
 \item{There is an increase in the number of non-degenerate levels by four. The 
doubly degenerate levels lying next to the non-degenerate levels of the last 
case become non-degenerate.}
 \item{The \textit{size} of the energy gap existing between the greatest 
negative level and 
smallest positive level falls from 0.0128 to 0.02.} 
 \end{itemize}

  The ground-state energy of this system is now obtained by summing the lowest 
fifty one levels and its value is -31.8219.  
  	
  	The upper half of the Fig. (\ref{D51}) gives the variation of spin 
density 
with the site index. The pattern obtained is seen to be similar to 
that of bond displacements except that the spin density maximum occurs at 
the 
site where the bond displacement is minimum and the oscillations within each 
envelope remain always above the value half. Also, the area under each of these 
envelopes, say from $n=35$ to $n=85$, is half and the sum of the areas under 
the 
two envelopes is equal to unity. This implies the extra ($51^{st}$) fermion 
distributes itself into these two envelopes with equal probability. Thus, the 
solitons with half 
integral spin are seen in the system. 

\subsection{For magnetization=49 (for $N_F$=99)}

 \begin{figure}[!h]
 \centerline{\psfig{figure=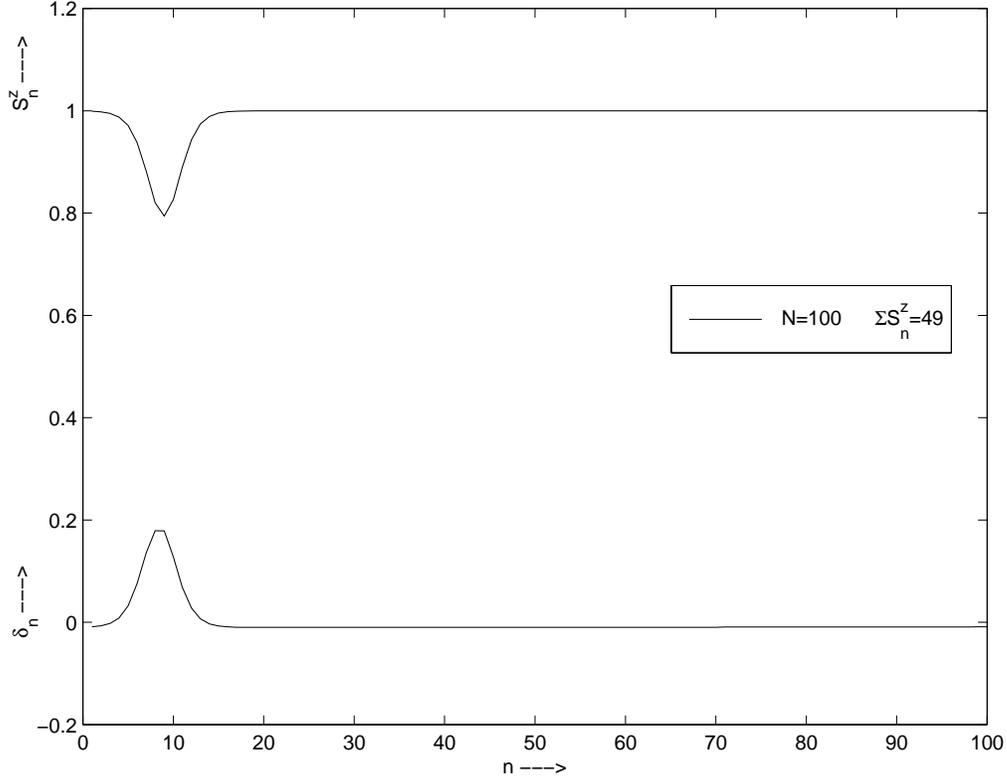,height=10.5 cm}}
 \caption{The bond displacements and expectation values of $S_n^z$ for 
magnetization=49.}
 \label{D99}
 \end{figure}

The bond displacements for the magnetization of the system being 49 or  
fermion filling being just one less than the maximum allowed, for the case of 
zero magnetic field are shown in the lower part of Fig. (\ref{D99}). It can 
be seen that the system has a single peak  ($\sim 0.18$) in bond 
displacements 
localized over a small region chosen randomly by the system by spontaneously 
breaking the translation symmetry.

  	The upper half of Fig. (\ref{D51}) gives the variation of spin 
density 
with the site index. It can be seen that there is a dip in the spin density 
over the region where there is a peak in bond displacements, showing that 
there is a localization of the missing ($100^{th}$) fermion over a small 
region. 

\subsection{For N=102, magnetization=18 (for $N_F$=69)} 

\begin{figure}[!h]
 \centerline{\psfig{figure=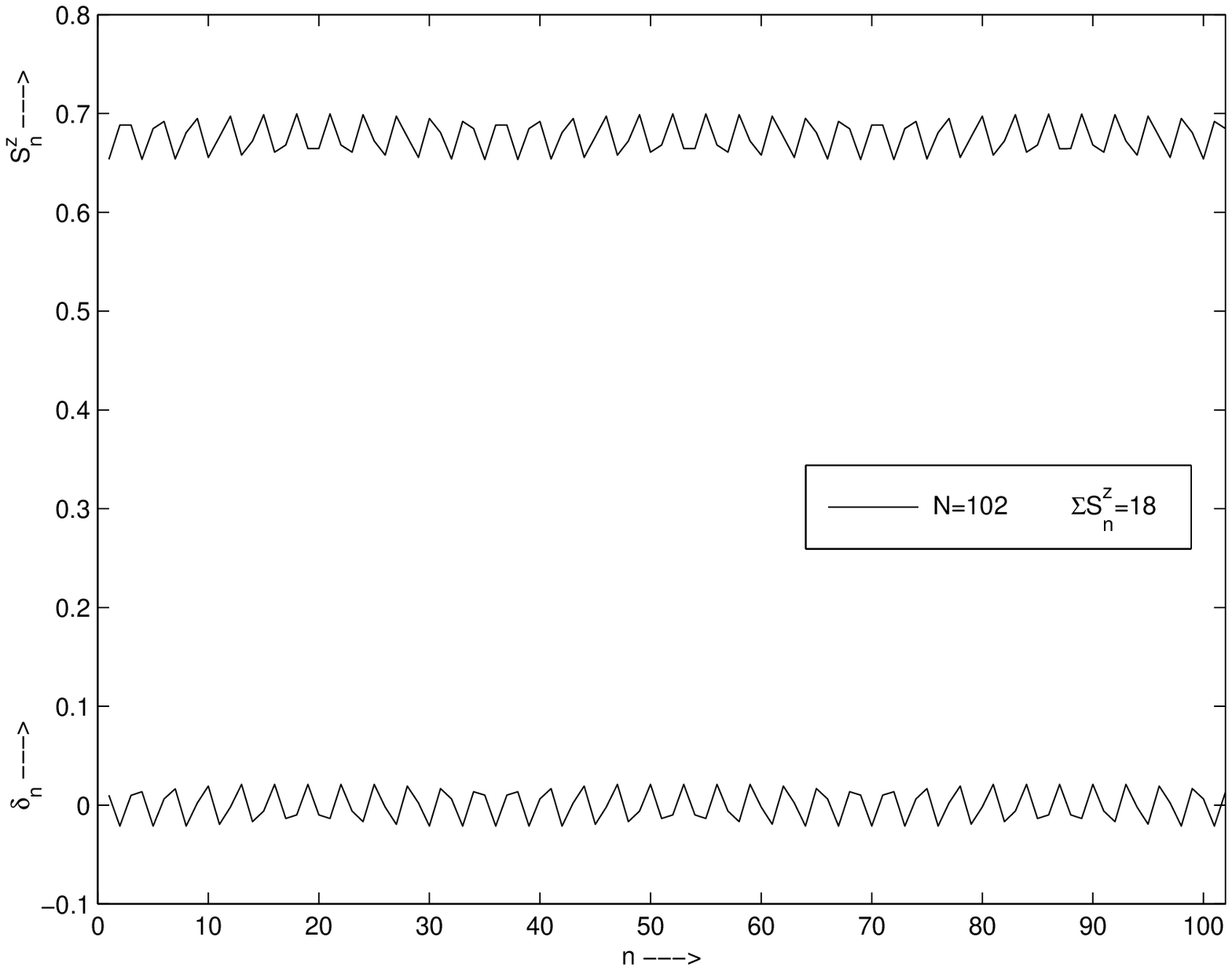,height=10.5 cm}}
 \caption{The bond displacements and expectation values of $S_n^z$ for 
magnetization=18.}
 \label{D69}
 \end{figure}

The bond displacements for a system size of 102 spins, $N_F$ being one 
away from trimerization i.e, 69, obtained by minimizing the Hamiltonian in Eq. 
(\ref{NFH}) for the case of zero magnetic field are shown in the lower part of  
Fig. (\ref{D69}). The bond displacements of the lattice spins are seen to 
oscillate around the value zero, their maximum magnitude being 0.022 and have 
an 
envelope whose periodicity is 34. The pattern is similar to $N_F=51$ case. 
Also, 
the single-particle energy spectrum is seen to be gapless in this case.
  	
  	The upper half of Fig. (\ref{D51}) gives the variation of spin 
density with the site index. The pattern obtained is similar to that of bond 
displacements except that the spin density maximum occurs at the site where the 
bond displacement is minimum and oscillations within each 
envelope remain always above the value 0.667. Also, the area under each of 
these 
envelopes, say from $n=24$ to $n=57$, is 0.333 and sum of the areas under all 
three envelopes is unity.  This means that the extra ( $69^{th}$) fermion 
distributes 
itself into the three envelopes with equal probability. Thus, solitons with 
fractional (1/3) spin are seen here in the system.

 \section{Variation of ground-state magnetization with magnetic field}
\begin{figure}[h]
 \centerline{\psfig{figure=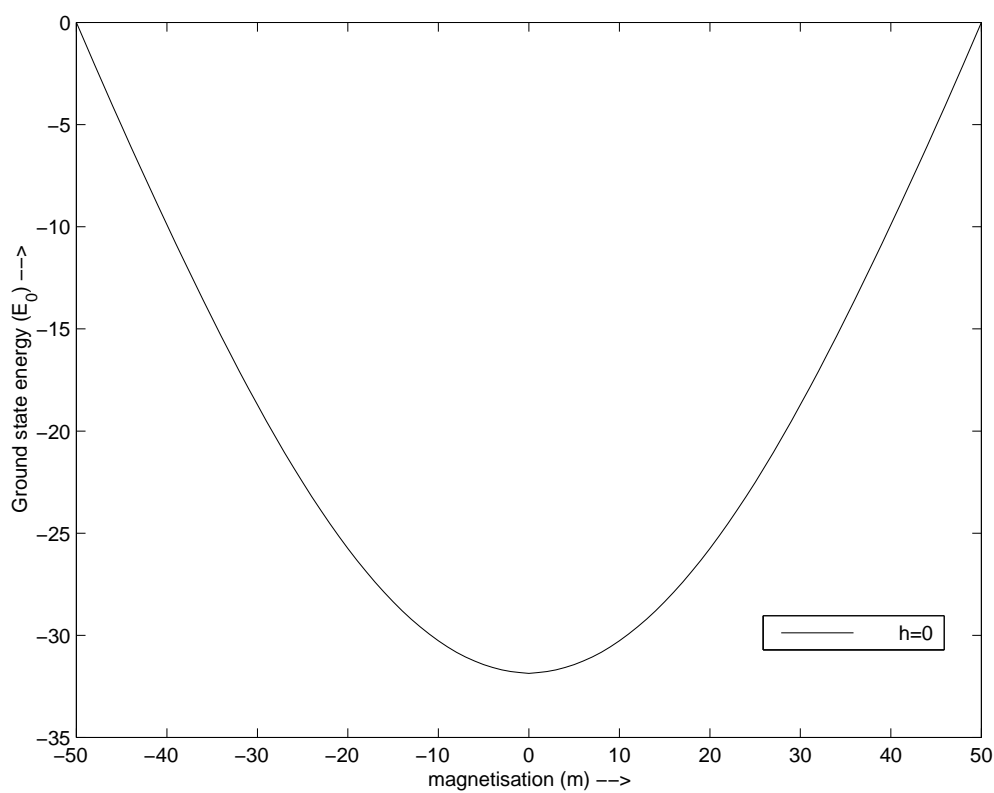,height=10.5 cm}}
 \caption{The variation of ground-state energy with magnetization $m$ for zero 
magnetic field $h$.} 
 \label{em}
 \end{figure} 
 
The single-particle energy spectrum, the ground-state energies and the spin 
densities at zero temperatures are calculated for each of the magnetizations (0 
to 50) possible in the system.  Fig. (\ref{em}) shows the variation of 
ground-state energy $E_0$ with magnetization $\textit{m}$, with the magnetic 
field term set to zero in the Hamiltonian (\ref{NFH}).

 \begin{figure}[h]
 \centerline{\psfig{figure=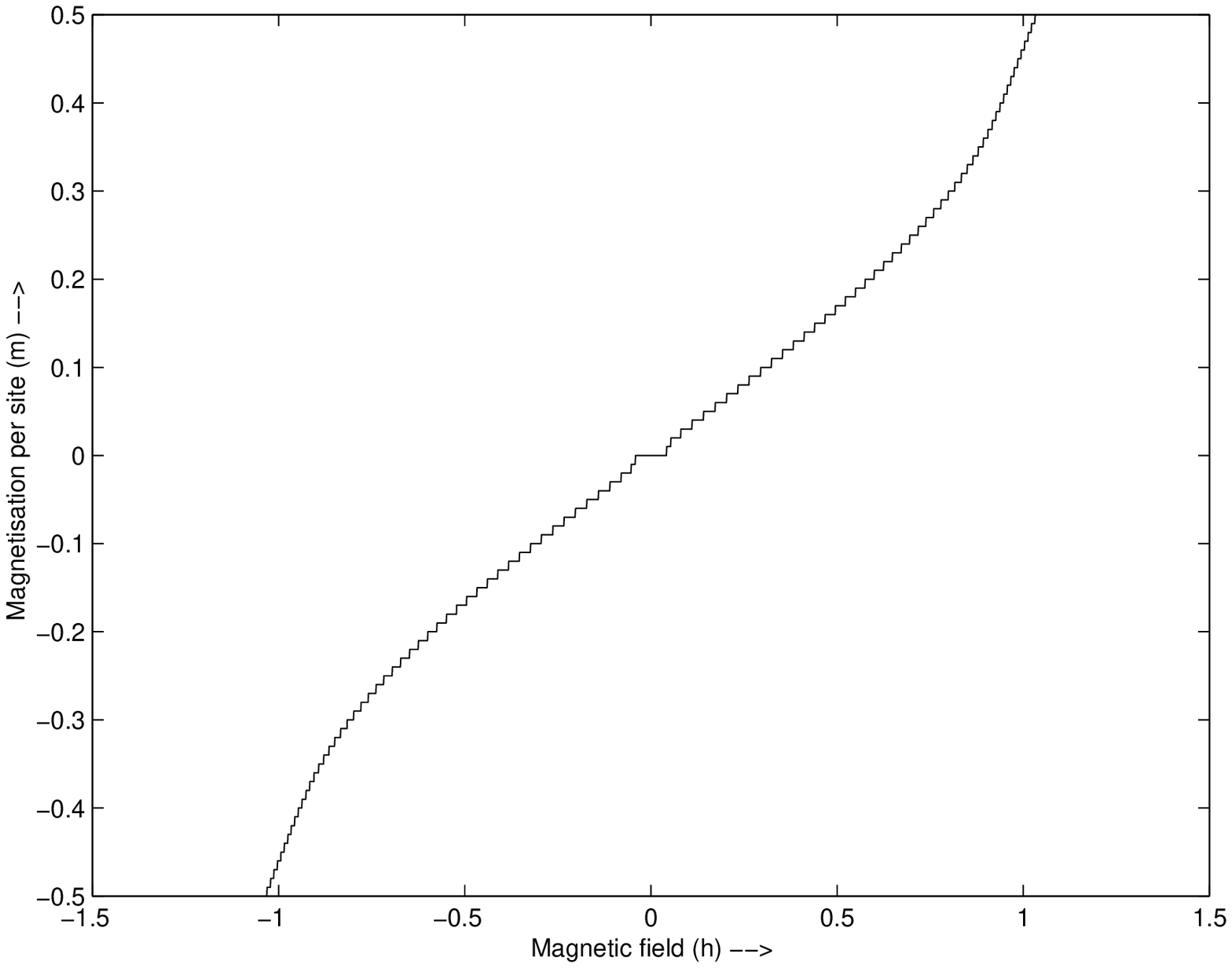,height=10.5 cm}}
 \caption{The variation of ground-state magnetization $\textit{m}$ with 
magnetic \mbox{field $\textit{h}$.}} 
 \label{mzm}
 \end{figure}

The variation of ground-state magnetization of the spin-Peierls system with 
magnetic field at zero temperature is now studied numerically, by varying the 
magnetic field $\textit{h}$ from  -1.1  to 1.1 in steps of 0.001 in the Zeeman 
term, $ h\sum_{n=1}^{N}[C_n^+C_n-\frac{1}{2}] $, of the Hamiltonian  
(\ref{NFH}). The ground-state energies for each of the possible magnetizations 
are 
calculated and the magnetization which minimizes the ground-state energy is 
identified for each value of the magnetic field.  The plot in Fig. (\ref{mzm}) 
shows the variation of  ground-state magnetization 
$\textit{m}$ with magnetic field $\textit{h}$. It is seen that the saturation 
magnetic field of this system is 1.035 and above this, the system exists only in 
the ferromagnetic state.

\section{Dimerized to uniform phase transition}

The dimerized to uniform structural phase transition in the spin-Peierls 
system 
shown in figure (\ref{PF}) is obtained numerically, for the case of zero 
magnetic 
field (\textit{h}=0) and zero magnetization ($\textit{m}=0$, $N_F$=50), as 
discussed below.

\begin{figure}[h]
 \centerline{\psfig{figure=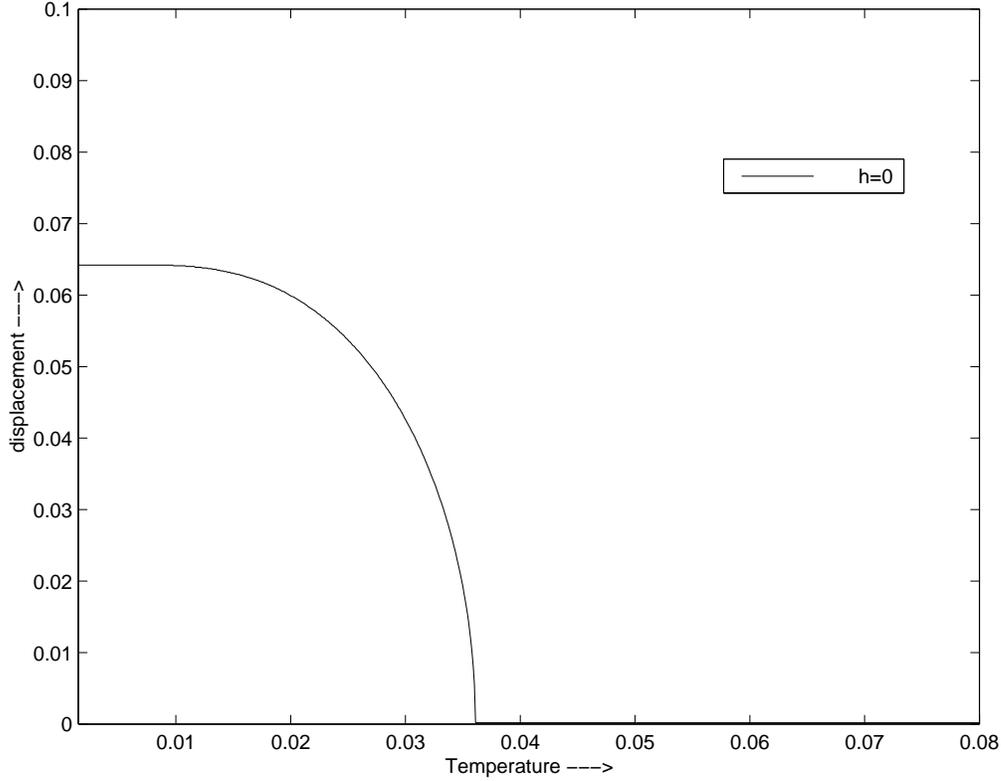,height=10.5 cm}}
 \caption{Dimerized to uniform phase transition at zero magnetic field.} 
 \label{NP}
 \end{figure} 

By taking the bond displacements to be of the form $\delta_n=\delta (-1)^n$, 
which 
is true for dimerized phase, the value of $\delta$ that minimizes the free 
energy is 
obtained for each non-zero temperature, by following the same procedure 
mentioned in section (2.2), except that free energy is minimized instead of the 
ground-state energy. The free 
energy \textit{U} at any temperature \textit{T}, using the Fermi-Dirac 
statistics, is given by 

\begin{eqnarray}
U=-\frac{1}{k_BT}\sum_{n=1}^{n=N}{\rm log}\left[1+{\rm exp}\right(\frac{E(n) 
-h}{k_B T}\left) \right]
\end{eqnarray}

\noindent where $E(n)$ are the single-particle energies at temperature 
\textit{T}, and $k_B$ is the Boltzmann constant \cite{7}.

The step sizes chosen in varying the temperature and the magnitude of the 
bond displacement while minimizing the free energy 
are both equal to  0.0001. Fig. (\ref{NP}) shows the second order phase 
transition, the critical parameter being the magnitude of the bond 
displacement, 
\textit{$\delta$}, plotted as a function of temperature, \textit{T}. The 
critical 
temperature as obtained from the graph is 0.036. It is to be noted 
that though the system is 1D in the Hamiltonian (\ref{NFH}) it still has a 
phase 
transition. This is because our calculation is equivalent to a mean-field 
theory 
which allows a non-zero displacement $\delta$. In a real $CuGeO_3$ system the 
interchain couplings justify this mean-field treatment and led to a finite 
temperature phase transition. 

\section{Calculation of the phonon frequencies }
\subsection{Analytical method}
In Section (2.2), single-particle energies were calculated assuming static 
lattice displacements. In this section, we will calculate the phonon 
frequencies using perturbation theory.  

The simplified spin-Peierls Hamiltonian is given by the Eq. (\ref{NFH}). The 
$i^{th}$ bond displacement, $\delta_{i+\frac{1}{2}}$, is given by 
\begin{center}
\begin{math}
\delta_{i+\frac{1}{2}}=-\lambda\left(u_{i+1} - u_{i} 
\right) \nonumber
\end{math}
\end{center}
\noindent

 Now the first order change in the ground-state energy (at zero 
temperature) 
$E_0$ due to the displacement of $i^{th}$ bond  proportional to
$\partial{E_0}/\partial{\delta_{i+\frac{1}{2}}}$. 
This is zero since $E_0$ is calculated by minimizing \textit{H} in Eq. 
(\ref{NFH}) with respect to all the bond displacements. Hence to 
realize the contribution of bond displacements to a change in the ground-
state energy, we have to find the derivative of second order with respect to the 
bond displacements.

Defining the elements of the dynamical matrix \textit{D} to be 

 \begin{center}
 \begin{math}
D_{{i+\frac{1}{2}},{j+\frac{1}{2}}}=\frac{\partial^2{E_0}}{\partial{\delta_{i+ 
\frac{1}{2}}}\partial{\delta_{j+\frac{1}{2}}}}
 \end{math}
 \end{center}

\noindent where all the derivatives are evaluated at those values of $\delta$'s 
which are obtained by minimizing the Hamiltonian (\ref{NFH}) \cite{6}. Using 
this 
definition for \textit{D} we get the lattice Hamiltonian to be 

\begin{equation}
H_{ph}=\frac{1}{2}M\sum_i \dot{u_i}^2+\frac{1}{2} 
\sum_{i,j}D_{i+\frac{1}{2}i,j+\frac{1}{2}}[\lambda(u_{i+1}-u_i)+\delta_{i+\frac
{1}{2}}^{(0)}][\lambda(u_{j+1}-u_j)+\delta_{j+\frac{1}{2}}^{(0)}]
\label{LH}
\end{equation}
\noindent where $u_i$ is the displacement of the ion at the
lattice point $i$, $M$ its mass, 
$\delta_{i+\frac{1}{2}}^{(0)}$ is the equilibrium value of 
$\delta_{i+\frac{1}{2}}$ which minimizes the Hamiltonian and 
$\lambda$ is the positive constant which is a measure of the coupling that 
exists between the lattice points and spins.

Now we define a transformation on displacements of lattice points as shown 
below,

 \begin{equation}
\tilde{u}_i=u_i-c_i 
 \label{UT}
 \end{equation}

\noindent we see that with periodic boundary conditions, and with the choice 
$c_1=0$ we 
have
 \begin{center}
\begin{math}
\lambda(c_{i+1}-c_i)+\delta_{i+\frac{1}{2}}^{(0)}=0
\end{math}
\end{center}
since  $\sum_i\delta_{i+\frac{1}{2}}=0$. 

Writing the lattice Hamiltonian (\ref{LH}) in terms of $\tilde{u}_i$'s, thus 
simplifies it to the form 
 
 \begin{eqnarray}
 H_{ph}&=&\frac{1}{2}M\sum_i 
\dot{\tilde{u}_i}^2+\frac{\lambda^2}{2}\sum_{i,j}D_{i,j}
 (\tilde{u}_{i+1}-\tilde{u}_i)(\tilde{u}_{j+1}-\tilde{u}_j) \nonumber
 \end{eqnarray}

\noindent Simplifying it further by defining 
 
\begin{equation}
\tilde{D}_{i,j}=D_{i-\frac{1}{2},j-\frac{1}{2}}+D_{i+\frac{1}{2},j+\frac{1}{2}}
-D_{i-\frac{1}{2},j+\frac{1}{2}}-D_{i+\frac{1}{2},j-\frac{1}{2}}
\end{equation}

\noindent we get the lattice Hamiltonian $H_{ph}$ (\ref{LH}) to be
 
\begin{eqnarray}
 H_{ph}&=&\frac{1}{2}M\sum_i 
\dot{\tilde{u}_i}^2+\frac{\lambda^2}{2}\sum_{i,j}\tilde{D}_{i,j}
 \tilde{u}_i\tilde{u}_j
 \label{MLH}
 \end{eqnarray}
 
 Now the frequency, $\omega_i$, of the $i^{th}$ phonon, (proportional 
to its energy) is given by

 \begin{equation}
 \omega_i=\lambda\sqrt{\frac{\mu_i}{M}}
 \label{freq}
 \end{equation}
 
\noindent where $\mu_i$ is the $i^{th}$ eigenvalue of the transformed dynamical 
matrix, 
$\tilde{D}$, and $M$ is the mass of each lattice point \cite{6}. 
 
 \begin{figure}[h]
 \centerline{\psfig{figure=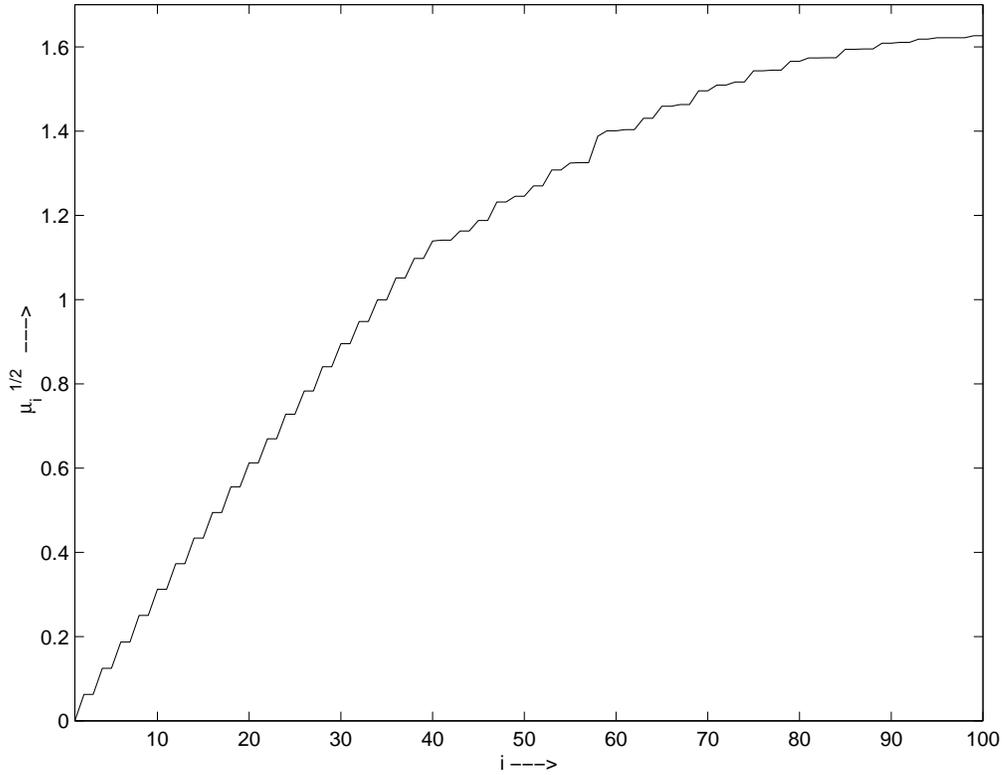,height=10.5 cm}}
 \caption{Square root of eigenvalues of the transformed dynamical matrix for 
the system in dimerized phase.}
 \label{phonon}
 \end{figure} 
 
 The transformed dynamical matrix for the dimerized phase is 
evaluated assuming bond displacements of the form 
  $\delta_i=\delta (-1)^i$
 and the root of the eigenvalues of the corresponding Hamiltonian are plotted in 
an increasing order as shown in Fig. (\ref{phonon}). This curve 
gives the phonon dispersion for the dimerized phase accurate upto a constant 
factor given 
by Eq. (\ref{freq}) \cite{8}.
While evaluating the 
elements of the transformed dynamical matrix, various symmetries like inversion 
symmetry about the bonds, translational symmetry of lattice points, etc., are 
used to generate all the elements from just the first few rows or columns.
  
 \subsection{Numerical method}
 The eigenfunctions of the Hamiltonian, $H$ for the case of dimerized phase 
(zero magnetization) are 
used to evaluate the transformed dynamical matrix using the second order 
perturbation theory as shown below

\begin{eqnarray}
&& \tilde{D}(m,n) \nonumber \\
&& =\sum_{j,k}\left[\frac{<n+1|H|n><n|H|n+1>+<m+1|H|m><m|H|m+1>}{E(j)-E(k)}
\right] \nonumber \\
&&
\end{eqnarray}

\noindent where N is the system size, j represents an occupied single-particle 
state and 
takes the values from 1 to 50, k represents an unoccupied single-particle state 
and ranges 
from 51 to 100, $E(j)$ and $E(k)$ denote the single-particle energies, $m$ and 
$n$ denote the site numbers with $|m>$ representing a state in which the 
$m^{th}$ site is occupied.
  While most of the matrix elements of $\tilde{D}$ obtained using the above 
equation are found to agree very well with those determined by the analytical 
method described in the previous section, a few off-diagonal elements are seen 
to differ slightly.
 
\chapter{Analytical Methods}

\section{Uniform phase}

\hspace{0.75in}The Hamiltonian for the system in uniform phase is given by 

\begin{eqnarray}
\textit{H}\hspace{0.2 cm}&=&\hspace{0.2 
cm}-\frac{J}{2}\sum_{n=1}^{N-1}\left[C_n^+
C_{n+1}+C_{n+1}^+C_n\right]\hspace{0.1 cm} 
+\hspace{0.1cm}\left(-1\right)^{N_F}\left[C_N^+C_{1}+C_{1}^+C_N 
\right]
\hspace{0.1 cm}\nonumber \\
& & \hspace{0.2 cm}+\hspace{0.1 
cm} 
\frac{c}{2}\sum_{n=1}^{N}\delta_{n+\frac{1}{2}}^2.
\label{UH}
\end {eqnarray}

\begin{figure}[h]
\centerline{\psfig{figure=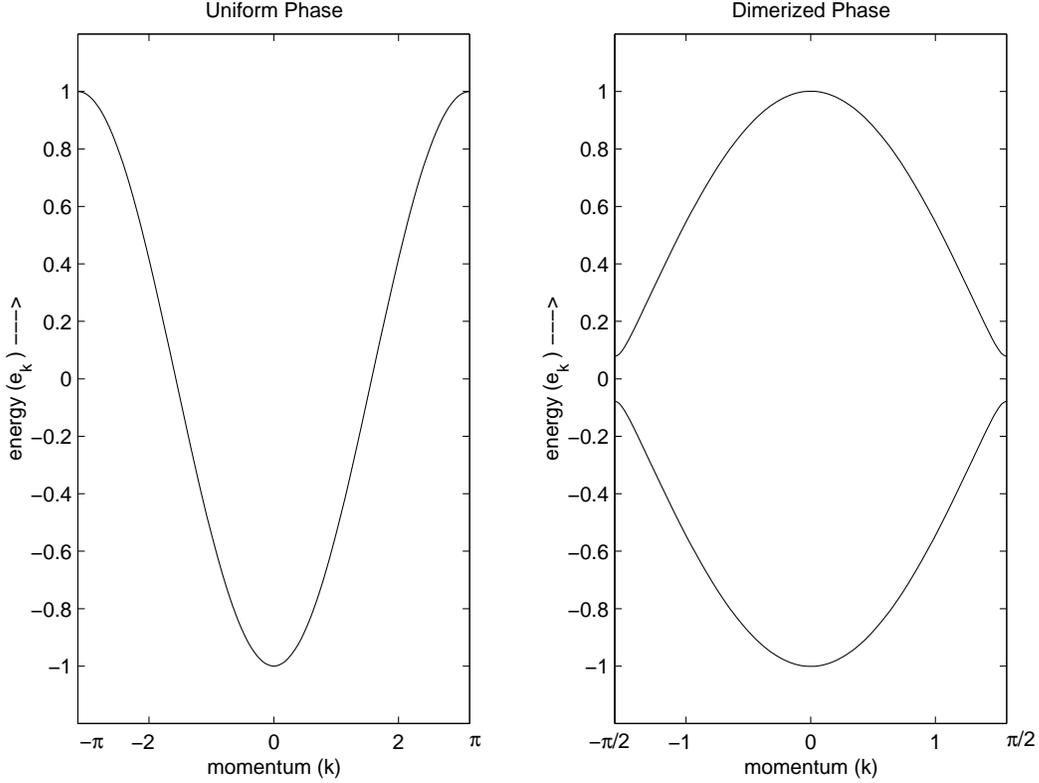,height=10.5 cm}}
\caption{The single-particle energy spectra for uniform and dimerised phases.}
\label{spe}
\end{figure}

This is obtained by setting each $\delta_{n+\frac{1}{2}}$ and also the total 
magnetization, m, to zero in (\ref{NFH}). The single-particle energy 
spectrum of this Hamiltonian is obtained by taking the discrete Fourier 
transform for this equation \cite{9,10}. This is given by $e_k = -J\cos(k)$, where k is the 
Fourier momentum and is shown in the left half of Fig. (\ref{spe}). In the plot, 
 the value of exchange integral $J$ is chosen to be unity. Also the lattice 
constant $l$ of uniform phase is set to unity all throughout this report. 
Imposing periodic boundary conditions on the system, $k$ assumes the form 
$k=2\pi n/N$ where $N$ is the system size and $n$ takes values from $0$ to 
$N-1$.

The ground-state energy of the system per site at zero temperature, $E_k$ is now 
obtained 
by summing the lower half of the energy spectrum and dividing it by the system 
size $N$, and is found to be $-J/\pi$. 

\section{Dimerized phase}
\hspace{0.75 in} In dimerized phase, the bond displacements are of the 
form $\delta_{n+\frac{1}{2}}=(-1)^n\delta$. The total magnetization of the 
system 
remains zero. The Hamiltonian of the system is shown in (\ref{DH}).

 \begin{eqnarray}
\textit{H}\hspace{0.2 cm}&=&\hspace{0.2 
cm}-\frac{J}{2}\sum_{n=1}^{N-1}\left[1+(-1)^n\delta\right]\left[C_n^+
C
_{n+
1}+C_{n+1}
^
+C_n\right]\hspace{0.1 cm} 
+\hspace{0.1cm}\left(-1\right)^{N_F}\left[C_N^+C_{1}+C_{1}^+C_N 
\right]
\hspace{0.1 cm}\nonumber \\
& & \hspace{0.2 cm}+\hspace{0.1 
cm} 
\frac{c}{2}\sum_{n=1}^{N}\delta^2.
\label{DH}
\end {eqnarray}

The Hamiltonian can be solved using the translation symmetry by redefining the 
size of unit cell in the dimerized phase to be twice that of the uniform phase 
and then applying the continuous Fourier transform as shown below. In equation 
(\ref{DN}), $A_k$, $B_k$ represents the 
annihilation operators for fermions with momentum $k$ associated with first and 
second of the pair of lattice points present in the unit 
cell respectively and their Hermitian conjugates represent the respective 
creation operators. 
 
 \begin{eqnarray}
 H&=&\int_{-\pi}^{\pi} 
\frac{dk}{\left(\frac{4\pi}{N}\right)}\left[A_k^+\hspace{0.2 cm} 
B_k^+\right] \left[\begin{array}{cc}
 \hspace{0.2 cm}0 \hspace{1 cm} -\frac{J+\delta}{2}-\frac{J-\delta}{2}e^{ik}\\
 -\frac{J+\delta}{2}-\frac{J-\delta}{2}e^{-ik}\hspace{0.6 cm} 0\hspace{0.2 cm}
 \end{array}\right] \left[\begin{array}{cc} A_k \\ B_k \end{array} \right]
 \label{DN}
 \end{eqnarray}
 
 \noindent The single-particle energy spectrum $e_k$, of this Hamiltonian is 
given by
 
  \begin{equation}
e_k=\pm{\sqrt \frac{J^2+\delta^2}{2}}\sqrt{1+\frac{J^2-\delta^2}{J^2+
\delta^2}\cos{k}}
  \end{equation}
 
 \noindent where $k$ goes from $-\pi$ to $\pi$ as shown in right half of Fig. 
(\ref{spe}). In the plot, exchange integral $J$ is set to unity.
 The ground-state energy of the system in this dimerized phase at zero 
temperature is calculated in the following section.
 
 \subsection{Ground-state energy}
 The total energy of the ground-state is obtained by summing all the negative 
energy states. From last section, it can be seen that the crystal momentum  $k$ 
takes values from $-\pi$ to $\pi$. Hence, the Ground-state energy per site, 
$E_d$ is given by

 \begin{eqnarray}
E_d&=&-\frac{Jb}{4\pi\sqrt2}\int_{-\pi}^\pi dk{\sqrt{1+a\cos{k}}} 
\nonumber \\ \nonumber \\
{\rm where}
 & & a=1-\frac{2\delta^2}{J^2} \nonumber \\
& &  b=1+\frac{\delta^2}{2J^2}
  \end{eqnarray}
  
\noindent This is an elliptical integral of the second kind. Using the series 
approximation for this integral and truncating the series to the quadratic 
power 
of $(\delta/J)$,

\begin{eqnarray}  
 E_d&=&-\frac{J}{\pi}\sqrt{1-\frac{\delta^2}{J^2}}\hspace {0.2 
cm}\epsilon(1-\frac{J}{\delta}) \nonumber \\
 &\approx&-\frac{J}{\pi}[1-\frac{\delta^2}{J^2}(\frac{1}{2}-a_1+2b_1
 {\rm log}({\frac{\delta}{J}}))] \nonumber \\ 
 &\approx&-\frac{J}{\pi}[1+\frac{\delta^2}{J^2}2b_1{\rm log}({\frac{J}{\delta}})] 
\nonumber \\ \nonumber \\    
{\rm where}
&&\epsilon(1-\frac{J}{\delta})=1+a_1(\frac{\delta^2}{J^2})-2b_1(\frac{\delta^2}
{
J^2}){\rm log}{(\frac{\delta}{J})} \nonumber \\ \vspace{0.1 cm}   {\rm and}
 & & a_1\approx 0.46301, \hspace{0.2cm} b_1\approx 0.24527
  \end{eqnarray} 
 
 Thus, the ground-state energy of the system in dimerized phase is lower 
compared to that of uniform phase for sufficiently smaller values of $\delta/J$ 
and hence dimerization occurs.
\section{Modeling the bond displacements for the case of 
magnetization=49 ($N_F=99$)}

\hspace{0.75 in}The Hamiltonian for the spin-Peierls system at zero temperature 
is shown in 
(\ref{NFH}). In the continuum limit, the eigenfunctions of this are of the form
\begin{math}
\psi_n=e^{in(\pi+k)}
\end{math}
\noindent where the lattice spacing is chosen to be unity and the magnetic 
field, 
$h$, is taken to be zero. Substituting this in the above Hamiltonian, we get 

\begin{eqnarray}
E_0 &=& \frac{1}{2}(1+\delta_{n+\frac{1}{2}})(e^{i(k+\pi)}+e^{-i(k+\pi)}) 
\nonumber \\
&  =&-(1+\delta_{n+\frac{1}{2}})\cos{k} \nonumber \\
& \approx& -1+v(x)+\frac{k^2}{2}
\end{eqnarray}

\noindent where $E_0$ is the energy eigenvalue corresponding to $\psi_n$ and 
$v(x)=-\delta_{n+\frac{1}{2}}$. It is to be noted that in the following 
discussion the lattice label $n$ is replaced by the continuum label $x$ 
whenever 
convenient. 

Using this form of $\delta_{n+\frac{1}{2}}$ in the Hamiltonian (\ref{NFH}) we 
get

\begin{eqnarray}
H&=&-1-\frac{1}{2}\partial^2_x+v(x)+\frac{c}{2}\int dx \hspace{0.2 cm}v^2(x)
\label{TH}
\end{eqnarray}
subject to the constraint $\int dx v(x)=0$. 

The minimum energy solution of this Hamiltonian, (\ref{TH}),is obtained using a 
trial function $\psi(x)=1/\cosh(\alpha x)$ where $\alpha >0$. 
Consequently, it  
reduces to 

\begin{eqnarray}
\frac{\tilde{e}_0}{\cosh({\alpha x})}&=&\frac{-1}{\cosh({\alpha  
x})}-\frac{1}{2}(\frac{\alpha^2}{\cosh({\alpha 
x})}-\frac{2\alpha^2}{\cosh^3({\alpha  
x})})+\frac{v(x)}{\cosh({\alpha x})} \nonumber \\
&&+\hspace{0.1 cm}(\frac{c}{2}\int dy \hspace{0.2 cm} v^2(y)\hspace{0.1 
cm})\frac{1}{\cosh({\alpha x})}
\end{eqnarray}

Hence we choose $v(x)=-\alpha^2/\cosh^2({\alpha x})$. Then matching the 
coefficient of $\psi(x)=1/\cosh(\alpha x)$ on the two sides gives the 
equation

 \begin{eqnarray}
\tilde{e}_0&=&-1-\frac{\alpha^2}{2}+\frac{c\alpha^3}{2}\int_{-\infty}^\infty 
\frac{dt}{\cosh^4{t}} \nonumber \\
\Rightarrow \tilde{e}_0&=&-1-\frac{\alpha^2}{2}+\frac{2c\alpha^3}{3}
\end{eqnarray}

\noindent The minimum energy of $\tilde{e}_0$ is attained when 
$\alpha=\frac{1}{2c}$ and it is equal to 
$-1-\frac{1}{24c^2}$ and the corresponding 
$v(x)=-\delta_{n+\frac{1}{2}}=-1/4c^2\cosh^2({\frac{x}{2c}})$.
For $c=1$, this value is equal to -1.125. From section (2.2.3), numerically 
this 
value is found to be -1.091. This discrepancy in the value of v(x) arises 
because the continuum approximation is valid only for $c>>1$. 
Also it can be noted that 

\begin{eqnarray}
\sum_n 
\delta_{n+\frac{1}{2}}&=&\int_{-\infty}^{\infty}\frac{dx}{4c^2\cosh^2({\frac{x}
{
2c
})}} 
\nonumber \\
&=&\frac{1}{c}
\end{eqnarray}

\noindent So in a finite system with N sites where $N>>c$, we introduce a 
Lagrangian multiplier in Eq. (\ref{TH} to enforce the constraint $dx \int 
v(x)=0$. We then find $\delta_{n+\frac{1}{2}}=-\frac{1}{Nc}$ for $n>>c$ instead 
of  $\delta_{n+\frac{1}{2}}=0$.Then 

\begin{equation}
\tilde{e}_0=-1-\frac{1}{24c^2}+\frac{1}{2Nc^2}. \nonumber
\end{equation}

\noindent It can be verified that this solution for $\psi(x)$ is indeed the 
minimum energy solution since the first 
order change in Hamiltonian (\ref{UH}) is zero. The analytical function of  
$\delta_{n+\frac{1}{2}}$ obtained here agrees perfectly well with the lower 
curve 
in Fig. (\ref{D99}) for the values $c=1$, $N=100$.

\chapter{ Scope for future work}
\section{Finite temperature calculations}
\hspace{0.75in} All the ground-state energies and eigenfunctions for the 
magnetizations discussed in section (2.2) are calculated at zero temperature. 
An 
attempt is made to extend these calculations using the same algorithm to 
non-zero temperatures using Fermi statistics as shown below. The free-energy  
$U$ at temperature T is now given by

\begin{eqnarray}
U=-\frac{1}{k_BT}\sum_{n=1}^{n=N}{\rm log}\left[1+{\rm exp}\right(\frac{E(n) 
-h}{k_B T}\left) \right]
\end{eqnarray}

\noindent where $E(n)$ are the single particle energies at temperature 
\textit{T} obtained after minimizing the Hamiltonian (\ref{NFH}) and $k_B$ is 
the Boltzmann constant \cite{7}. However, it is noticed that convergence in finding the 
set of lattice displacements which minimize the free-energy could not be 
achieved. Also, the same problem was encountered while calculating the single 
particle energies and eigenfunctions for low-lying excited states at zero 
temperature. A Solution is yet to be found to overcome this problem. This is 
required to obtain the complete phase diagram and compare it with Fig. 
(\ref{PF}).

\section{Calculation of change in eigenfunctions due to dynamical phonons}
In section (2.2), a procedure has been detailed to evaluate all the 
eigenfunctions assuming the lattice points to be static. The phonon frequencies 
are then calculated in section (2.5) to leading order in lattice displacements 
by using perturbation theory. The corrections to the unperturbed eigenfunctions 
due to these phonons can thereafter be obtained self-consistently. Work to 
accomplish this is in progress. This will give a more accurate picture of the 
soliton profile in the presence of dynamical phonons. For instance, the soliton 
profiles shown in Figs. (\ref{D51}) and (\ref{D69}) which do not take phonons 
into account now get modified.  

\section{Modeling the lattice spin displacements for all possible 
magnetizations}
The lattice spin displacements for all possible magnetizations can be modelled 
by adopting a procedure similar to the one discussed in section (3.3). A trial 
function for the lattice displacements is used to minimize the Hamiltonian for 
that particular magnetization. The necessary constraints required to be 
satisfied are incorporated in this equation using Lagrangian multipliers. By 
comparing the analytical shapes of the bond displacements with those observed 
numerically, one can find the value of the lattice-spin coupling parameter 
$\lambda$.

\section{Inclusion of the $S_n^z S_{n+1}^z$ term in the Hamiltonian}
As detailed in section (2.2.1), for all calculations in this report the 
generalised Heisenberg spins in the full Hamiltonian (\ref{SH}) are 
replaced by the XY spins. This is done to simplify the calculations without 
qualitatively affecting the phase diagram. However, a more realistic picture  
requires the inclusion of the spin-spin interactions in the $\hat{z}$ direction 
which on Jordan-Wignerization become four fermion interactions \cite{4,5}. 
Techniques like the density matrix renormailsation group and bosonization 
can be employed to 
\nopagebreak
\enlargethispage*{1000pt}
study the Hamiltonian of such a system \cite{3}.

\clearpage
\bibliography{References}

\thispagestyle{empty}
\end{document}